\documentclass[11pt,a4paper]{article}
\usepackage[T1]{fontenc}
\usepackage{setspace}
\usepackage[latin1]{inputenc}
\usepackage{amsmath}
\usepackage{amssymb}
\usepackage{setspace}
\usepackage{esint}

\def\adot{{\dot{\alpha}}}
\def\bdot{{\dot{\beta}}}

\def\btheta{{\bar{\theta}}}
\def\blambda{{\bar{\lambda}}}
\def\bD{{\bar{D}}}
\def\bxi{{\bar{\xi}}}

\def\ahat{{{\hat{\alpha}}}}
\def\bhat{{{\hat{\beta}}}}
\def\ghat{{{\hat{\gamma}}}}

\def\ap{{\alpha^\prime}}

\def\be{\begin{equation}}
\def\ee{\end{equation}}
\def\ba{\begin{eqnarray}}
\def\ea{\end{eqnarray}}

\date{}

\allowdisplaybreaks

\numberwithin{equation}{section}

\title{On the $\mathcal{N}$ = 4, d = 4 pure spinor measure factor}

\author{Thales Azevedo\thanks{thales@ift.unesp.br%
}}

\begin{document}

\maketitle
\begin{center}
ICTP South American Institute for Fundamental Research \\
Instituto de F\'isica Te\'orica, UNESP - Univ. Estadual Paulista \\
Rua Dr. Bento T. Ferraz 271, 01140-070, S\~ao Paulo, SP, Brasil.
\par\end{center}

\

\begin{abstract}In this work, we obtain a simple measure factor for the $\lambda$ and $\theta$ zero-mode integrations in the pure-spinor formalism in the context of an $\mathcal{N}$ = 4, d = 4 theory. We show that the measure can be defined unambiguously up to BRST-trivial terms and an overall factor, and is much simpler than (although equivalent to) the expression obtained by dimensional reduction from the $\mathcal{N}$ = 1, d = 10 measure factor. We also give two explicit examples of how to obtain the dual to a vertex operator using this measure.\end{abstract}

\section{Introduction}

\

The prescription for computing tree-level open superstring scattering
amplitudes in a manifestly super-Poincar\'e covariant manner was given by Berkovits some time ago, in the same paper in which the pure spinor superstring was introduced \cite{nathan}. In this formalism, the unintegrated vertex operators are in the ghost-number 1 cohomology of the BRST operator
\be
Q=\frac{1}{2\pi\mathrm{i}}\oint {\mathrm{d}} z\,\lambda^\ahat d_\ahat
\label{BRST}
\ee
and are functions of the ten-dimensional superspace coordinates $x^\mu$ and $\theta^\ahat$ for $\mu = 0$ to 9 and $\ahat = 1$ to 16. More information on notations and conventions can be found in Appendix \ref{App}.

In (\ref{BRST}), $\lambda^\ahat$ is a pure-spinor ghost variable, i.e. $\lambda^\ahat$ satisfies $\lambda\gamma_\mu\lambda=0$, and $d_\ahat = p_\ahat - \frac{2}{\ap}\left[\partial x^\mu (\theta\gamma_\mu)_\ahat +\frac{1}{2} (\theta\gamma^\mu\partial\theta)(\theta\gamma_\mu)_\ahat\right]$, where $p_\ahat$ is the conjugate momentum to $\theta^\ahat$. The OPE's $$p_\ahat(z)\,\theta^\bhat(w) \sim \frac{\delta_\ahat^\bhat}{z-w} \qquad \textrm{and} \qquad x^\mu(z,\bar{z})\,x^\nu(w,\bar{w}) \sim -\frac{\ap}{2}\eta^{\mu\nu}\log |z-w|^2$$ imply
\be
d_\ahat(z)\,{\mathcal{F}}(x(w),\theta(w)) \sim \frac{1}{z-w} D_\ahat{\mathcal{F}}(x(w),\theta(w))\,,
\ee
where $D_\ahat = \frac{\partial}{\partial\theta^\ahat} + (\theta\gamma^\mu)_\ahat \partial_\mu$ and ${\mathcal{F}}(x,\theta)$ is any superfield. Hence, we write $Q{\mathcal{F}} = \lambda^\ahat D_\ahat{\mathcal{F}}$.

In addition to knowing the vertex operators and OPE's, one needs to know  how to perform the integrations of the $\lambda$ and $\theta$ zero modes. They  can be performed by means of the following BRST-invariant measure factor:
\be
\Big\langle (\lambda \gamma^\mu \theta)(\lambda \gamma^\nu \theta)(\lambda \gamma^\rho \theta)(\theta \gamma_{\mu\nu\rho} \theta)\Big\rangle = 1\,.
\label{flat}
\ee
More precisely, the integration is done by keeping only the terms proportional to three $\lambda$'s and five $\theta$'s in this combination.

Thus far, we have written everything in an ${\mathcal{N}}$ = 1, d = 10 notation.
In order to compute superstring scattering amplitudes in an $\mathcal{N}$ = 4, d = 4 theory --- such as the gauge theory describing the effective world-volume degrees of freedom of a D3-brane, for instance --- using the pure-spinor formalism, one needs to know how to perform the integrations of the $\lambda$ and $\theta$ zero modes in that case. In other words, one needs to find a BRST-invariant measure factor analogous to (\ref{flat}).

At first, it might seem to be just a matter of dimensional reduction.
However, although the particular combination of $\lambda$'s and $\theta$'s of (\ref{flat}) is special in  ten flat dimensions, since it is the unique (up to an overall factor) SO($9,1$) scalar which can be built out of three $\lambda$'s and five $\theta$'s, there is no reason why its dimensional reduction should be preferred over any other BRST-invariant, SO($3,1$)$\times$SU(4) scalar  in four dimensions. Therefore, it is important to investigate whether there is any ambiguity in the definition of the $\mathcal{N}$ = 4, d~=~4 measure factor.

In this paper, this issue is studied in detail. In section 2, we write the most general four-dimensional expression with three $\lambda$'s and five $\theta$'s and derive the conditions for it to be BRST invariant. In section 3, we find the independent BRST-trivial combinations of the terms introduced in section 2. In section 4, we present the main results of this paper.
 We find that the $\mathcal{N}$~=~4, d = 4 measure factor is unique up to BRST-trivial terms and an overall factor. Moreover, we show that the measure can be written in a much simpler form than the dimensional reduction of (\ref{flat}) --- the latter has twelve terms, whereas the former has only three. In section 5, we give two examples of the use of this measure factor. Finally, section 6 is devoted to our conclusions.

\section{BRST equations\label{sec:review}}

\

In four-dimensional notation, the most general SO($3,1$)$\times$SU(4)-invariant, real expression one can write with three $\lambda$'s and five $\theta$'s is
\ba
 (\lambda^3\theta^5) \!\!\! & := & \!\!\! c_1\,\varepsilon_{mnj\ell}(\blambda_i\blambda_k)(\lambda^m\theta^n)(\theta^i\theta^j)(\theta^k\theta^\ell) \nonumber \\
 \!\!\! &  & \!\!\! +\;c_2\, (\blambda_j\blambda_k)(\blambda_\ell\btheta_i)(\theta^i\theta^j)(\theta^k\theta^\ell) + c_3\,\varepsilon_{mnj\ell}(\blambda_i\btheta_k)(\lambda^k\theta^\ell)(\lambda^m\theta^n)(\theta^i\theta^j) \nonumber \\
 \!\!\! &  & \!\!\! +\;c_4\, (\blambda_\ell\blambda_k)(\lambda^j\theta^\ell)(\theta^i\theta^k)(\btheta_i\btheta_j) + c_5\,(\blambda_i\btheta_k)(\blambda_j\btheta_\ell)(\lambda^k\theta^\ell)(\theta^i\theta^j)  \label{fourdim}\\
 \!\!\! &  & \!\!\! +\;c_6\, \varepsilon_{mn\ell k}(\lambda^i\theta^k)(\lambda^j\theta^\ell)(\lambda^m\theta^n)(\btheta_i\btheta_j)  \nonumber\\
\!\!\! &  & \!\!\! +\;\mathrm{H.c.}\,,\nonumber
\ea
where $c_1, \ldots, c_6$ are arbitrary constants  and ``H.c.'' means ``Hermitian conjugate''. We use the standard d=4 two-component spinor notation as described in Appendix \ref{twocomp}. In particular, $i,j,\ldots = 1$ to 4 and $\alpha,\adot = 1$ to 2. One can convince oneself  these are the only non-zero independent terms which can be constructed, keeping in mind that
\be
\lambda^{\alpha i}\blambda^\adot_i = 0 \qquad \mathrm{and} \qquad
(\lambda^i\lambda^j)=\frac{1}{2}\varepsilon^{ijk\ell}(\blambda_k\blambda_\ell)\,,
\label{pure}
\ee
which are the dimensional reduction of $\lambda \gamma^\mu \lambda=0$. More details on dimensional reduction can be found in Appendix \ref{dimred}.
The notation we are going to use throughout the paper is such that
\be
(\lambda^3\theta^5) =: \sum_{n=1}^6 c_n \mathbf{T}_n + \mathrm{H.c.}\,,
\ee
i.e. we define $\mathbf{T}_1, \ldots, \mathbf{T}_6$ to be the independent possible terms as appearing in (\ref{fourdim}). For example, $\mathbf{T}_3 \equiv \varepsilon_{mnj\ell}(\blambda_i\btheta_k)(\lambda^k\theta^\ell)(\lambda^m\theta^n)(\theta^i\theta^j)$. The $\mathbf{T}_n$ and their Hermitian conjugates $\mathbf{T}^\dagger_n$ form a basis for four-dimensional expressions made of three $\lambda$'s and five $\theta$'s.  For example, it is not difficult to show the dimensional reduction of (\ref{flat}) gives (\ref{fourdim}) with $c_1=1$, $c_2=c_4=4$, $c_3=3$, $c_5=12$ and $c_6=2$, up to an overall factor.

Since we are looking for the pure spinor measure, we are interested in expressions which are annihilated by $\lambda^\ahat D_\ahat = \lambda^{\alpha p} D_{\alpha p} + \blambda_{\adot p} \bD^{\adot p}$. This requirement yields equations for the constants in (\ref{fourdim}). We begin with
\be
{\left[\lambda^\ahat D_\ahat (\lambda^3\theta^5)\right]}_{\theta^4\btheta^0} = \lambda^{\alpha p} D_{\alpha p} [c_1\mathbf{T}_1] + \blambda_{\adot p} \bD^{\adot p} [c_2\mathbf{T}_2 + c_3\mathbf{T}_3]\,,
\ee
where the subscript $\theta^4\btheta^0$ means ``contributions with four $\theta$'s and no $\btheta$.'' The explicit calculation gives:
\ba
\lambda^{\alpha p} D_{\alpha p} \mathbf{T}_1  \!\!\! & = & \!\!\!  \varepsilon_{mnj\ell}(\blambda_i\blambda_k)\,\Big[\,(\lambda^m\lambda^n)(\theta^i\theta^j)(\theta^k\theta^\ell) + (\lambda^m\theta^n)(\theta^i\lambda^j)(\theta^k\theta^\ell)\nonumber\\
 \!\!\! &  & \!\!\!    \qquad \qquad  \qquad \qquad \qquad  \qquad \qquad \quad \!\!\!+\; (\lambda^m\theta^n)(\theta^i\theta^j)(\theta^k\lambda^\ell) \,\Big] \nonumber\\
\!\!\! & = & \!\!\! 4\, (\blambda_i\blambda_k)(\blambda_j\blambda_\ell)(\theta^i\theta^j)(\theta^k\theta^\ell) \,,
\ea
where we used (\ref{pure}) and $\lambda^{\alpha[i|}\lambda^{\beta|j]} = -\frac{1}{2}\varepsilon^{\alpha\beta}(\lambda^i\lambda^j)$. Moreover,
\be
\blambda_{\adot p}\bD^{\adot p}\mathbf{T}_2 =  (\blambda_j\blambda_k)(\blambda_\ell\blambda_i)(\theta^i\theta^j)(\theta^k\theta^\ell)
\ee
and $\blambda_{\adot p}\bD^{\adot p}\mathbf{T}_3 = 0$. Therefore
\be
\boxed{{\left[\lambda^\ahat D_\ahat (\lambda^3\theta^5)\right]}_{\theta^4\btheta^0} = 0 \;\Longleftrightarrow c_2 = 4c_1\,.}
\label{eq1}
\ee

Proceeding to the next order, we have
\be
{\left[\lambda^\ahat D_\ahat (\lambda^3\theta^5)\right]}_{\theta^3\btheta^1} = \lambda^{\alpha p} D_{\alpha p} [c_2\mathbf{T}_2 + c_3\mathbf{T}_3] + \blambda_{\adot p} \bD^{\adot p} [c_4\mathbf{T}_4 + c_5\mathbf{T}_5 + c_6\mathbf{T}_6]\,.
\ee
The $\mathbf{T}_2$-contribution is easy to compute. We get
\be
\lambda^{\alpha p} D_{\alpha p} \mathbf{T}_2 = - (\blambda_j\blambda_k)(\blambda_\ell\btheta_i)(\lambda^i\theta^j)(\theta^k\theta^\ell)\,.
\ee
For $\mathbf{T}_3$, we obtain
\ba
\lambda^{\alpha p} D_{\alpha p} \mathbf{T}_3  \!\!\! & = & \!\!\!  -\varepsilon_{mnj\ell}(\blambda_i\btheta_k)\,\Big[\,(\lambda^k\lambda^\ell)(\lambda^m\theta^n)(\theta^i\theta^j) - (\lambda^k\theta^\ell)(\lambda^m\lambda^n)(\theta^i\theta^j)\nonumber\\
 \!\!\! &  & \!\!\!    \qquad \qquad  \qquad \qquad \qquad  \qquad \qquad \quad \,\, -\, (\lambda^k\theta^\ell)(\lambda^m\theta^n)(\theta^i\lambda^j) \,\Big] \nonumber\\
\!\!\! & = & \!\!\! 4\, (\blambda_j\blambda_\ell)(\blambda_i\btheta_k)(\lambda^k\theta^\ell)(\theta^i\theta^j) \,.
\ea
The $\mathbf{T}_4$- and $\mathbf{T}_5$-contributions are also simple to calculate:
\ba
\blambda_{\adot p} \bD^{\adot p} \mathbf{T}_4  \!\!\! & = & \!\!\! - (\blambda_\ell\blambda_k)(\lambda^j\theta^\ell)(\theta^i\theta^k)(\blambda_i\btheta_j) \,, \\
\blambda_{\adot p} \bD^{\adot p} \mathbf{T}_5  \!\!\! & = & \!\!\! - (\blambda_i\btheta_k)(\blambda_j\blambda_\ell)(\lambda^k\theta^\ell)(\theta^i\theta^j)\,.
\ea
Finally, $\blambda_{\adot p} \bD^{\adot p} \mathbf{T}_6 = 0$. In all, we get our second equation for the coefficients:
\be
\boxed{{\left[\lambda^\ahat D_\ahat (\lambda^3\theta^5)\right]}_{\theta^3\btheta^1} = 0 \;\Longleftrightarrow c_2 + 4c_3 = c_4 + c_5\,.}
\label{eq2}
\ee

We now analyze the contributions with equal number of $\theta$'s and $\btheta$'s:
\be
{\left[\lambda^\ahat D_\ahat (\lambda^3\theta^5)\right]}_{\theta^2\btheta^2} = \lambda^{\alpha p} D_{\alpha p} [c_4\mathbf{T}_4 + c_5\mathbf{T}_5 + c_6\mathbf{T}_6] + \blambda_{\adot p} \bD^{\adot p} [\bar{c}_4\mathbf{T}_4^\dagger + \bar{c}_5\mathbf{T}_5^\dagger + \bar{c}_6\mathbf{T}_6^\dagger]\,.
\ee
Again, it is straightforward to compute  the contributions from $\mathbf{T}_4$ and $\mathbf{T}_5$:
\ba
\lambda^{\alpha p} D_{\alpha p} \mathbf{T}_4  \!\!\! & = & \!\!\! - (\blambda_\ell\blambda_k)(\lambda^j\theta^\ell)(\lambda^i\theta^k)(\btheta_i\btheta_j) \,, \\
\lambda^{\alpha p} D_{\alpha p} \mathbf{T}_5  \!\!\! & = & \!\!\! (\blambda_i\btheta_k)(\blambda_j\btheta_\ell)(\lambda^k\lambda^\ell)(\theta^i\theta^j) \,.
\ea
The $\mathbf{T}_6$-contribution yields
\ba
\lambda^{\alpha p} D_{\alpha p} \mathbf{T}_6  \!\!\! & = & \!\!\!  \varepsilon_{mn\ell k}(\btheta_i\btheta_j)\,\Big[\,(\lambda^i\lambda^k)(\lambda^j\theta^\ell)(\lambda^m\theta^n) - (\lambda^i\theta^k)(\lambda^j\lambda^\ell)(\lambda^m\theta^n)\nonumber\\
 \!\!\! &  & \!\!\!   \qquad \qquad  \qquad \qquad \qquad  \qquad \qquad \;\, +\; (\lambda^i\theta^k)(\lambda^j\theta^\ell)(\lambda^m\lambda^n) \,\Big] \nonumber\\
\!\!\! & = & \!\!\! 4\, (\blambda_\ell\blambda_k)(\lambda^i\theta^k)(\lambda^j\theta^\ell)(\btheta_i\btheta_j) \,.
\ea
These in turn imply, by Hermitian conjugation,
\ba
\blambda_{\adot p} \bD^{\adot p} \mathbf{T}_4^\dagger  \!\!\! & = & \!\!\! - (\lambda^\ell\lambda^k)(\blambda_j\btheta_\ell)(\blambda_i\btheta_k)(\theta^i\theta^j) \,, \\
\blambda_{\adot p} \bD^{\adot p} \mathbf{T}_5^\dagger  \!\!\! & = & \!\!\! (\lambda^i\theta^k)(\lambda^j\theta^\ell)(\blambda_k\blambda_\ell)(\btheta_i\btheta_j) \,, \\
\blambda_{\adot p} \bD^{\adot p} \mathbf{T}_6^\dagger  \!\!\! & = & \!\!\! 4\, (\lambda^\ell\lambda^k)(\blambda_i\btheta_k)(\blambda_j\btheta_\ell)(\theta^i\theta^j) \,.
\ea
Thus we obtain our last equation:
\be
\boxed{{\left[\lambda^\ahat D_\ahat (\lambda^3\theta^5)\right]}_{\theta^2\btheta^2} = 0 \;\Longleftrightarrow \bar{c}_5 = c_4 + 4c_6\,,}
\ee
as well as its complex conjugate.

Note that the vanishing of the orders $\theta^1\btheta^3$ and $\theta^0\btheta^4$  implies the complex conjugates of (\ref{eq1}) and (\ref{eq2}), since they are just the Hermitian conjugates of the orders $\theta^3\btheta^1$ and $\theta^4\btheta^0$, respectively.

In summary, we have the following system of equations:
\be
\lambda^\ahat D_\ahat (\lambda^3\theta^5) = 0 \; \Longleftrightarrow \; \left\{ \begin{array}{rcl} c_2 \!\!\! & = & \!\!\! 4c_1 \\
 c_2 + 4c_3 \!\!\! & = & \!\!\! c_4 + c_5 \\
\bar{c}_5 \!\!\! & = & \!\!\! c_4 + 4c_6 \end{array} \right. ,
\label{system}
\ee
as well as their complex conjugates.

\section{BRST-trivial combinations}

\

In the last section, we found the equations which the constants in (\ref{fourdim}) have to satisfy for the expression to be BRST-invariant. Because there are less equations than constants, one might think the $\mathcal{N}$ = 4, d = 4 pure spinor measure factor is then not unambiguously defined. Fortunately, that is not the case, and the seemingly independent expressions are actually related by BRST-trivial terms, as we show in the following.

In order to find the independent BRST-trivial  combinations of the $\mathbf{T}_n$, i.e. $\!$the combinations which equal $\lambda^\ahat D_\ahat$ of something, we start by looking for all independent possible terms with two $\lambda$'s and six $\theta$'s. Keeping (\ref{pure}) in mind, we find that there are five:
\begin{subequations}\label{triviais}
\ba
\chi_1 \!\!\! & := & \!\!\! (\lambda^i\theta^j)(\theta^k\theta^\ell)(\blambda_k\btheta_\ell)(\btheta_i\btheta_j)\,,\\
\chi_2 \!\!\! & := & \!\!\! \varepsilon_{ijk\ell}(\lambda^i\theta^j)(\lambda^m\theta^k)(\theta^\ell\theta^n)(\btheta_m\btheta_n)\,,\\
\chi_3 \!\!\! & := & \!\!\! \varepsilon^{ijk\ell}(\blambda_i\btheta_j)(\blambda_m\btheta_k)(\btheta_\ell\btheta_n)(\theta^m\theta^n)\,,\\
\chi_4 \!\!\! & := & \!\!\! \varepsilon_{mnj\ell}(\blambda_i\btheta_k)(\lambda^m\theta^n)(\theta^i\theta^j)(\theta^k\theta^\ell)\,,\\
\chi_5 \!\!\! & := & \!\!\! \varepsilon^{mnj\ell}(\lambda^i\theta^k)(\blambda_m\btheta_n)(\btheta_i\btheta_j)(\btheta_k\btheta_\ell)\,.
\ea
\end{subequations}

 Acting with $\lambda^\ahat D_\ahat= \lambda^{\alpha p} D_{\alpha p} + \blambda_{\adot p} \bD^{\adot p}$ on these terms, we obtain BRST-trivial expressions made of three $\lambda$'s and five $\theta$'s. We begin with the first one:
\ba
\lambda^{\alpha p} D_{\alpha p} \chi_1  \!\!\! &=& \!\!\!   (\lambda^i\theta^j)(\theta^k\lambda^\ell)(\blambda_k\btheta_\ell)(\btheta_i\btheta_j) \nonumber\\
\!\!\! &=& \!\!\! \frac{1}{2}\left[\mathbf{T}_4^\dagger - \mathbf{T}_5^\dagger\right],
\ea
\ba
\blambda_{\adot p} \bD^{\adot p}  \chi_1  \!\!\! &=& \!\!\! -(\lambda^i\theta^j)(\theta^k\theta^\ell)(\blambda_k\btheta_\ell)(\btheta_i\blambda_j)\nonumber\\
\!\!\! &=& \!\!\! \frac{1}{2}\left[\mathbf{T}_5 - \mathbf{T}_4\right].
\ea
Thus we find the first BRST-trivial expression:
\be
\boxed{
 \mathbf{T}_4^\dagger - \mathbf{T}_5^\dagger + \mathbf{T}_5 - \mathbf{T}_4 = \lambda^\ahat D_\ahat \left[2\chi_1\right].}
\ee
Of course, we could multiply the expression on the left-hand side of this equation  by any constant and it would remain BRST-trivial. The same applies to the other boxed expressions we find in the following.

For the second term in (\ref{triviais}), we have
\ba
\lambda^{\alpha p} D_{\alpha p} \chi_2  \!\!\! &=& \!\!\! \varepsilon_{ijk\ell}(\btheta_m\btheta_n) \Big[ (\lambda^i\lambda^j)(\lambda^m\theta^k)(\theta^\ell\theta^n) - (\lambda^i\theta^j)(\lambda^m\lambda^k)(\theta^\ell\theta^n) \nonumber \\
\!\!\!  &  &  \!\!\!  \qquad\qquad\qquad +\; (\lambda^i\theta^j)(\lambda^m\theta^k)(\lambda^\ell\theta^n) - (\lambda^i\theta^j)(\lambda^m\theta^k)(\theta^\ell\lambda^n) \Big] \nonumber\\
  \!\!\! &=& \!\!\! 4\mathbf{T}_4 - \mathbf{T}_6\,,
\ea
\ba
\blambda_{\adot p} \bD^{\adot p}  \chi_2  \!\!\! & = & \!\!\! -\varepsilon_{ijk\ell}(\lambda^i\theta^j)(\lambda^m\theta^k)(\theta^\ell\theta^n)(\btheta_m\blambda_n) \nonumber \\ 
\!\!\! & = & \!\!\!  \mathbf{T}_3\,. 
\ea
Therefore,
\be
\boxed{
 \mathbf{T}_3 + 4\mathbf{T}_4 - \mathbf{T}_6 = \lambda^\ahat D_\ahat \chi_2\,.}
\ee

Since the third term in (\ref{triviais}) is equal to $\chi_2^\dagger$,
\be
\boxed{
\mathbf{T}_3^\dagger + 4\mathbf{T}_4^\dagger - \mathbf{T}_6^\dagger = \lambda^\ahat D_\ahat \chi_3\,.}
\ee

For the fourth term,

\ba
\lambda^{\alpha p} D_{\alpha p} \chi_4  \!\!\! &=& \!\!\! -\varepsilon_{mnj\ell}(\blambda_i\btheta_k) \Big[ (\lambda^m\lambda^n)(\theta^i\theta^j)(\theta^k\theta^\ell) + (\lambda^m\theta^n)(\theta^i\lambda^j)(\theta^k\theta^\ell) \nonumber \\
\!\!\!  &  &  \!\!\!  \qquad\qquad\qquad\; -\, (\lambda^m\theta^n)(\theta^i\theta^j)(\lambda^k\theta^\ell) + (\lambda^m\theta^n)(\theta^i\theta^j)(\theta^k\lambda^\ell) \Big] \nonumber\\
  \!\!\! &=& \!\!\! 4\mathbf{T}_2 - \mathbf{T}_3\,,
\ea
\ba
\blambda_{\adot p} \bD^{\adot p}  \chi_4  \!\!\! &=& \!\!\! \varepsilon_{mnj\ell}(\blambda_i\blambda_k)(\lambda^m\theta^n)(\theta^i\theta^j)(\theta^k\theta^\ell)\nonumber\\
\!\!\! &=& \!\!\! \mathbf{T}_1\,.
\ea
Therefore,
\be
\boxed{
\mathbf{T}_1 + 4\mathbf{T}_2 -\mathbf{T}_3 = \lambda^\ahat D_\ahat \chi_4\,.}
\ee

Finally, the last term in (\ref{triviais}) is equal to $\chi_4^\dagger$, so
\be
\boxed{
\mathbf{T}_1^\dagger + 4\mathbf{T}_2^\dagger -\mathbf{T}_3^\dagger = \lambda^\ahat D_\ahat \chi_5\,.}
\ee

\section{The $\mathcal{N}$ = 4, d = 4 measure factor in a simple form\label{sec:cghost}}

\

We are now in position to show the $\mathcal{N}$ = 4, d = 4 measure factor is unique up to BRST-trivial terms and an overall factor.
Consider once again the most general real expression with three $\lambda$'s and five $\theta$'s of (\ref{fourdim}). One has
\be
(\lambda^3\theta^5) = c_1\mathbf{T}_1 + c_2\mathbf{T}_2 + c_3\mathbf{T}_3 + c_4\mathbf{T}_4 + c_5\mathbf{T}_5 + c_6\mathbf{T}_6 + \mathrm{H.c.}\,.
\ee
If this is BRST-invariant, then the constants satisfy the equations (\ref{system}) and their complex conjugates.
We are free to add BRST-trivial terms to the above expression. If  we add
$$
-c_1\left[\mathbf{T}_1 + 4\mathbf{T}_2 - \mathbf{T}_3\right]  -\frac{1}{4}(c_4+c_5)\left[\mathbf{T}_3 + 4\mathbf{T}_4 -\mathbf{T}_6\right] + \mathrm{H.c.}
$$
to $(\lambda^3\theta^5)$, we get\footnote{Note the equal sign here means ``equal up to BRST-trivial terms.''}
\be
(\lambda^3\theta^5) = -c_5\mathbf{T}_4 + c_5\mathbf{T}_5 + {1 \over 4}( c_5 +\bar{c}_5) \mathbf{T}_6 + \mathrm{H.c.}\,,
\ee
where we used (\ref{system}).

Furthermore, if $c_5 = \alpha + \mathrm{i}\beta$, with $\alpha, \beta \in \mathbb{R}$, then we can add the BRST-trivial term
$$
-\mathrm{i} \beta \left[ \mathbf{T}_5 - \mathbf{T}_5^\dagger - \mathbf{T}_4 + \mathbf{T}_4^\dagger\right]
$$
to $(\lambda^3\theta^5)$, thus obtaining
\be
\boxed{
(\lambda^3\theta^5) = -\alpha\left[\mathbf{T}_4 - \mathbf{T}_5 - {1 \over 2} \mathbf{T}_6 + \mathrm{H.c.}\right].
}
\ee
This shows that the measure is unique up to BRST-trivial terms and an overall factor.

The measure can be even further simplified, provided that we relax its manifest reality condition. By adding the BRST-trivial terms
\be
\alpha \left[\mathbf{T}_4^\dagger - \mathbf{T}_5^\dagger +\mathbf{T}_5 - \mathbf{T}_4\right] + \frac{1}{2}\alpha\left[\mathbf{T}_3 + 4\mathbf{T}_4 - \mathbf{T}_6\right],
\ee
we arrive at
\be
(\lambda^3\theta^5) = \frac{1}{2} \alpha\left[\mathbf{T}_3 + 4\mathbf{T}_5 + \mathbf{T}_6^\dagger\right],
\ee
or, more explicitly,
\be
\boxed{
\begin{array}{cl}
\Big\langle \varepsilon_{mnj\ell}(\blambda_i\btheta_k)(\lambda^k\theta^\ell)(\lambda^m\theta^n)(\theta^i\theta^j) 
\!\!\!\! &  +\; 4\,(\blambda_i\btheta_k)(\blambda_j\btheta_\ell)(\lambda^k\theta^\ell)(\theta^i\theta^j)   \\
&  +\; \varepsilon^{mn\ell k}(\blambda_i\btheta_k)(\blambda_j\btheta_\ell)(\blambda_m\btheta_n)(\theta^i\theta^j)\Big\rangle = \displaystyle\frac{1}{240}\,,
\end{array}
} 
\label{measure}
\ee
where we recovered the explicit form of each term and used the normalization of (\ref{flat}). 

This simple $\mathcal{N}$ = 4, d = 4 pure spinor measure factor is the main result of this work. Note that, while the dimensional reduction of (\ref{flat}) yields twelve independent terms, this expression has only three.
 In the next section, we give two explicit examples of how to obtain the dual to a vertex operator using this measure factor.

\section{Examples}

\subsection{Gluino}

\

Consider the following vertex operator from \cite{fleury}:
\be
\widetilde{V}_{\mathrm{gluino}} = \frac{1}{4}(\lambda\gamma^\mu\theta)(\lambda\gamma^\nu\theta)(\theta\gamma_{\mu\nu}\xi^*)\,,
\ee
where $\xi_\ahat^*$ is the zero-momentum gluino antifield.\footnote{We choose to work at zero-momentum for the sake of simplicity, and also because only in this computation is the explicit form of the measure factor needed. Momentum corrections at higher $\theta$-orders can in principle be obtained from the ten-dimensional expressions by carrying out the same steps as in this section (dimensionally reducing, simplifying by adding BRST-trivial terms and then looking for the dual to the vertex operator), but beyond zero-momentum one just has to impose that the terms with more than five $\theta$'s in the amplitude cancel.} It is easy to show this operator is annihilated by $Q$. The dimensional reduction yields
\ba
\widetilde{V}_{\mathrm{gluino}} \!\!\! & = & \!\!\! 2\,(\blambda_j\blambda_k)(\theta^j\theta^\ell)(\theta^k\xi_\ell^*) + 3\, (\blambda_i\btheta_j)(\theta^i\theta^k)(\blambda_k\bxi^{*j})  - 2\, (\blambda_j\btheta_i)(\theta^i\theta^k)(\blambda_k\bxi^{*j})         \nonumber \\
\!\!\! &  & \!\!\! -\,6\,(\blambda_i\btheta_j)(\lambda^j\theta^k)(\theta^i\xi^*_k)   -\varepsilon_{ijk\ell}(\lambda^i\theta^j)(\lambda^n\theta^\ell)(\theta^k\xi^*_n)   +4\,(\blambda_{j}\btheta_{i})(\lambda^k\theta^i)(\theta^j\xi^*_k)               \nonumber\\
\!\!\! &  & \!\!\! -\,2\,(\blambda_{i}\btheta_{j})(\lambda^k\theta^i)(\theta^j\xi^*_k)   + (\blambda_i\blambda_j)(\theta^k\theta^i)(\btheta_k\bxi^{*j}) -3\,\varepsilon_{ijk\ell}(\lambda^m\theta^i)(\lambda^k\theta^\ell)(\btheta_m\bxi^{*j})  \nonumber\\
\!\!\! &  & \!\!\!  +\;2\,\varepsilon_{ijk\ell} (\lambda^i\theta^j)(\theta^\ell\theta^m)(\blambda_m\bxi^{*k})\nonumber\\
\!\!\! &  & \!\!\! +\;\mathrm{H.c.}\,.
\label{V}
\ea

We can simplify this expression by adding BRST-trivial terms. For example,\begin{subequations}
\ba
Q\left[\varepsilon_{ijk\ell}(\lambda^i\theta^j)(\theta^n\theta^\ell)(\theta^k\xi^*_n)\right] \!\!\! & = & \!\!\!
 -4\,(\blambda_\ell\blambda_k)(\theta^n\theta^\ell)(\theta^k\xi^*_n)-\varepsilon_{ijk\ell}\,(\lambda^i\theta^j)(\lambda^n\theta^\ell)(\theta^k\xi^*_n) \nonumber\\
\!\!\! & = & \!\!\! -4\mathbf{t}_1 - \mathbf{t}_5\,, 
\ea
where $\mathbf{t}_N$ refers to the $N$-th term in $\widetilde{V}_{\mathrm{gluino}}$ as appearing in (\ref{V}), without the numerical factor (e.g. $\mathbf{t}_1 \equiv \,(\blambda_j\blambda_k)(\theta^j\theta^\ell)(\theta^k\xi_\ell^*))$. So the combination $4\mathbf{t}_1 + \mathbf{t}_5$ is BRST-trivial. Of course, this implies $4\mathbf{t}_1^\dagger + \mathbf{t}_5^\dagger$ is also BRST-trivial.

One can also show
\be
Q\left[(\blambda_i\btheta_j)(\theta^j\theta^k)(\theta^i\xi^*_k)\right] = -\mathbf{t}_1  - \mathbf{t}_4  - \mathbf{t}_6 \,,
\ee

\be
Q\left[ (\btheta_i\btheta_j)(\lambda^j\theta^k)(\theta^i\xi^*_k)\right] = \mathbf{t}_4 + \mathbf{t}_2^\dagger - \mathbf{t}_8^\dagger\,,
\ee

\be
Q\left[ (\btheta_i\btheta_j)(\lambda^k\theta^i)(\theta^j\xi^*_k)\right] = \mathbf{t}_6 + \mathbf{t}_7 + \mathbf{t}_3^\dagger + \mathbf{t}_8^\dagger\,,
\ee

\be
Q\left[ \varepsilon_{ijk\ell} (\lambda^i\theta^j)(\theta^\ell\theta^m)(\btheta_m\bxi^{*k})\right] = -3\mathbf{t}_8 - \mathbf{t}_9 - \mathbf{t}_{10}\,,
\ee
\end{subequations}
as well as their Hermitian conjugates.
Then we can add the BRST-trivial amount 
\be
-2\mathbf{t}_1 - 3\mathbf{t}_2 +2\mathbf{t}_3 - 9\mathbf{t}_4
+\mathbf{t}_5 - 4\mathbf{t}_6 + 2\mathbf{t}_7 - \mathbf{t}_8
- 2\mathbf{t}_9 - 2\mathbf{t}_{10} + \mathrm{H.c.}
\ee
to $\widetilde{V}_{\mathrm{gluino}}$ to obtain a BRST-equivalent vertex operator given by
\be
\widetilde{V}_{\mathrm{gluino}}^\prime = -15\, (\blambda_i\btheta_j)(\lambda^j\theta^k)(\theta^i\xi^*_k) - 5\,\varepsilon_{ijk\ell}(\lambda^m\theta^i)(\lambda^k\theta^\ell)(\btheta_m\bxi^{*j}) + \mathrm{H.c.}\,.
\ee
Finally, multiplying by an overall factor and dropping the prime, we arrive at the simplest form
\be\boxed{
\widetilde{V}_{\mathrm{gluino}} = 3\, (\blambda_i\btheta_j)(\lambda^j\theta^k)(\theta^i\xi^*_k) + \varepsilon_{ijk\ell}(\lambda^m\theta^i)(\lambda^k\theta^\ell)(\btheta_m\bxi^{*j}) + \mathrm{H.c.}\,.}
\label{Vfinal}
\ee

Now we may look for the dual to this vertex operator, meaning the BRST-closed expression with one $\lambda$ and two $\theta$'s whose product with $\widetilde{V}_{\mathrm{gluino}}$ gives something proportional to the measure (\ref{measure}). The dual is certainly going to contain the gluino field $\xi^\ahat$, and the substitution
\be
\xi_\bhat^*\xi^\ahat \longrightarrow \delta_\bhat^\ahat
\label{relation}
\ee
can be used to determine it.

Comparing (\ref{Vfinal}) with (\ref{measure}), we expect the dual to contain a term with $\varepsilon_{ijk\ell}$. It is easy to see there is only one such term:
\be
V_{\mathrm{gluino}}^{[1]} = \kappa_1\,\varepsilon_{ijk\ell}(\lambda^i\theta^j)(\theta^k\xi^\ell)\,,
\ee
where $\kappa_1$ is a constant to be determined. Then, using (\ref{relation}), we obtain
\be
\widetilde{V}_{\mathrm{gluino}}\,V_{\mathrm{gluino}}^{[1]} = \kappa_1\,\left[3\mathbf{T}_3 + 3\mathbf{T}_5\right].
\ee

To obtain a term proportional to $\mathbf{T}_6^\dagger$, we need a term with $\blambda$. There are not many, and it is not difficult to show the following one works:
\be
V_{\mathrm{gluino}}^{[2]} = \kappa_2\,(\blambda_i\btheta_j)(\theta^i\xi^j)\,,
\ee
for a constant $\kappa_2$ to be determined shortly. Making use of (\ref{relation}) once again, we get
\be
\widetilde{V}_{\mathrm{gluino}}\,V_{\mathrm{gluino}}^{[2]} = -\kappa_2\,\left[3\mathbf{T}_5 + \mathbf{T}_6^\dagger \right].
\ee

Hence, if $\kappa_1 = \frac{1}{3}$ and $\kappa_2 = -1$, we have
\be
\widetilde{V}_{\mathrm{gluino}}\,V_{\mathrm{gluino}} := \widetilde{V}_{\mathrm{gluino}} (V_{\mathrm{gluino}}^{[1]} + V_{\mathrm{gluino}}^{[2]}) =\mathbf{T}_3 + 4\mathbf{T}_5 + \mathbf{T}_6^\dagger\,,
\ee
which means the dual to $\widetilde{V}_{\mathrm{gluino}}$ is given by
\be
\boxed{
V_{\mathrm{gluino}} = \frac{1}{3}\varepsilon_{ijk\ell}(\lambda^i\theta^j)(\theta^k\xi^\ell) - (\blambda_i\btheta_j)(\theta^i\xi^j)\,.
}
\ee
One can show this expression is BRST-invariant, as it should be.

\subsection{Gluon}

\

We now derive, for the gluon, expressions analogous to those obtained for the gluino in the previous subsection. The vertex operator for the zero-momentum gluon antifiled $a^*_b$ ($b=0$ to 3) can also be found in \cite{fleury}, and is given by
\be
\widetilde{V}_{\mathrm{gluon}} = \frac{1}{4}(\lambda\gamma^\mu\theta)(\lambda\gamma^\nu\theta)(\theta\gamma_{\mu\nu}{}^b\theta)a^*_b\,.
\ee
Again, it is easy to show this operator is annihilated by $Q$. The dimensional reduction can be easily obtained from the gluino case, by noting that $\xi^* \longmapsto (\gamma^b\theta)a^*_b \; \Longrightarrow \; \widetilde{V}_{\mathrm{gluino}} \longmapsto \widetilde{V}_{\mathrm{gluon}}$. One gets
\ba
\widetilde{V}_{\mathrm{gluon}} \!\!\! & = & \!\!\! 3\,(\blambda_j\blambda_k)(\theta^j\theta^\ell)(\theta^k a^* \btheta_\ell) - 3\, (\blambda_i\btheta_j)(\theta^i\theta^k)(\theta^{j}a^*\blambda_k)  + 2\, (\blambda_j\btheta_i)(\theta^i\theta^k)(\theta^{j}a^*\blambda_k)         \nonumber \\
\!\!\! &  & \!\!\! -\,8\,(\blambda_i\btheta_j)(\lambda^j\theta^k)(\theta^i a^*\btheta_k)   -4\,\varepsilon_{ijk\ell}(\lambda^i\theta^j)(\lambda^n\theta^\ell)(\theta^k a^*\btheta_n)   +4\,(\blambda_{j}\btheta_{i})(\lambda^k\theta^i)(\theta^j a^*\btheta_k)               \nonumber\\
\!\!\! &  & \!\!\!  -\,2\,\varepsilon_{ijk\ell} (\lambda^i\theta^j)(\theta^\ell\theta^m)(\theta^{k} a^*\blambda_m)\nonumber\\
\!\!\! &  & \!\!\! +\;\mathrm{H.c.}\,,
\label{V2}
\ea
where $(\theta^i a^* \blambda_j) = \theta^{\alpha i} a^*_{\alpha\adot}\blambda_j^\adot$, with $a^*_{\alpha\adot}=(\sigma^b)_{\alpha\adot}a^*_b$.

To simplify this expression, one can add combinations of the following BRST-trivial terms:\begin{subequations}
\be
Q\left[\varepsilon_{ijk\ell}(\lambda^i\theta^j)(\theta^n\theta^\ell)(\theta^k a^*\btheta_n)\right]  =  -4\mathbf{u}_1 - \mathbf{u}_5 +\mathbf{u}_7\,, 
\ee

\be
Q\left[(\blambda_i\btheta_j)(\theta^j\theta^k)(\theta^i a^*\btheta_k)\right] = -\mathbf{u}_1 +\mathbf{u}_3  - \mathbf{u}_4  - \mathbf{u}_6 \,,
\ee

\be
Q\left[ (\theta^j\theta^k)(\btheta_k\btheta_j)(\theta^{i} a^*\blambda_i)\right] = -2\left(\mathbf{u}_2 + \mathbf{u}_3 + \mathbf{u}_4 +\mathbf{u}_6\right),
\ee
\end{subequations}
as well as their Hermitian conjugates, where $\mathbf{u}_N$ refers to the $N$-th term in $\widetilde{V}_{\mathrm{gluon}}$ as appearing in (\ref{V2}), without the numerical factor (e.g. $\mathbf{u}_1 \equiv \,(\blambda_j\blambda_k)(\theta^j\theta^\ell)(\theta^k a^*\btheta_\ell))$. Then one can show the following vertex operator is equivalent to $\widetilde{V}_{\mathrm{gluon}}$:
\be
\widetilde{V}_{\mathrm{gluon}}^\prime = -6\mathbf{u}_5 +12\mathbf{u}_6 + \mathrm{H.c.}\,.
\ee
Finally, multiplying by an overall factor and dropping the prime, we arrive at the simplest form
\be\boxed{
 \widetilde{V}_{\mathrm{gluon}} = \varepsilon_{ijk\ell}(\lambda^i\theta^j)(\lambda^n\theta^\ell)(\theta^k a^*\btheta_n)   -2\,(\blambda_{j}\btheta_{i})(\lambda^k\theta^i)(\theta^j a^*\btheta_k) + \mathrm{H.c.}\,. } 
\label{V2final}
\ee

The dual to this vertex operator is a  BRST-closed expression with one $\blambda$ and one $\theta$ whose product with $\widetilde{V}_{\mathrm{gluon}}$ gives something proportional to the measure (\ref{measure}). This expression should also contain the gluon field $a_b$. It is easy to see there is only one possibility:
\be
\boxed{
V_{\mathrm{gluon}} = (\theta^{i} a\, \blambda_i)\,.
}
\label{thing}
\ee
Then, using $a^*_b a_c \longrightarrow \eta_{bc}$, it is not difficult to show $\widetilde{V}_{\mathrm{gluon}}V_{\mathrm{gluon}}$ is indeed proportional to (\ref{measure}).

%\footnote{In fact, one could also choose the Hermitian conjugate of (\ref{thing}). Then the product $\widetilde{V}V$ would be proportional to the Hermitian conjugate of the measure (\ref{measure}).}

\section{Conclusion}

\

In this work, we have obtained a simple measure factor for the $\lambda$ and $\theta$ zero-mode integrations in the pure-spinor formalism in the context of an $\mathcal{N}$ = 4, d = 4 theory. We have shown that the measure can be defined unambiguously up to BRST-trivial terms and an overall factor, and is much simpler than (although equivalent to) the expression obtained by dimensional reduction from the $\mathcal{N}$ = 1, d = 10 measure factor. We have also given two explicit examples of how to obtain the dual to a vertex operator using this measure.

We expect these results to be useful for the computation of  disk scattering amplitudes of states propagating in the world-volume of a D3-brane, as well as open-closed superstring amplitudes of states
close to the AdS${}_5$ boundary \cite{azevedo}. 
It would also be interesting to make contact with recent results obtained for ten-dimensional superstring amplitudes, such as those in \cite{stieberger,mafra,mafra&schlotterer}.

% Of course, future applications which we have not thought about could also be possible.

\

\textbf{Acknowledgements:} I would like to thank Thiago Fleury, Renann Jusinskas and my advisor Nathan
Berkovits
for useful discussions and suggestions. This work was supported by
FAPESP grants 2010/19596-2 and 11/11973-4.

%\newpage

\appendix

\section{Appendix}\label{App}

\subsection{Two-component spinor notation}\label{twocomp}

\

The four-dimensional Lorentz group SO$(3,1)$ is locally isomorphic to ${\mathrm{SL(2,}\,\mathbb{C})}$, which has two distinct fundamental representations. One of them is described by a pair of complex numbers \cite{wess}
\be
\psi_\alpha = \left(\begin{array}{c} \psi_1 \cr \psi_2 \end{array}\right)\,,
\ee
with transformation law
\be
\psi_\alpha^\prime = \Lambda_\alpha^\beta \psi_\beta\,, \qquad \Lambda \in {\mathrm{SL(2,}\,\mathbb{C})}\,,
\label{rep1}
\ee
and is called $\left({1 \over 2},0\right)$ or left-handed chiral representation.

The other fundamental representation, called $\left(0,{1 \over 2}\right)$ or right-handed chiral, is obtained by complex conjugation:
\be
\bar{\psi}_\adot^\prime = \bar{\Lambda}_\adot^\bdot \bar{\psi}_\bdot \,, \qquad \bar{\Lambda}_\adot^\bdot = \overline{(\Lambda_\alpha^\beta)}\,.
\label{rep2}
\ee
The dot over the indices indicates the representation to which we refer.

The indices with and without dot are raised and lowered in the following way:
\begin{subequations}
\be
\psi^\alpha = \varepsilon^{\alpha\beta} \psi_\beta\,, \qquad \bar{\chi}^\adot = \varepsilon^{\adot\bdot}\bar{\chi}_\bdot\,;
\ee
\be
\psi_\alpha = \varepsilon_{\alpha\beta} \psi^\beta\,, \qquad \bar{\chi}_\adot = \varepsilon_{\adot\bdot}\bar{\chi}^\bdot\,,
\ee
\end{subequations}
where $\varepsilon$ is antisymmetric and has the properties
\be
\varepsilon^{12} = \varepsilon^{\dot{1}\dot{2}} = -\varepsilon_{12} = -\varepsilon_{\dot{1}\dot{2}} = 1\;\;\; \Longrightarrow \;\;\; \varepsilon_{\alpha\beta}\varepsilon^{\beta\gamma} = \delta_{\alpha}^{\gamma}\,, \qquad \varepsilon_{\adot\bdot}\varepsilon^{\bdot\dot{\gamma}} = \delta_{\adot}^{\dot{\gamma}}\,.
\ee
For spinorial derivatives, raising or lowering the indices involve an extra sign. For example, $D^\alpha_i = -\varepsilon^{\alpha\beta} D_{\beta i}$.

The convention for contraction of spinorial indices is
\be
\psi^\alpha \lambda_\alpha =: (\psi\lambda)\,, \qquad \bar{\chi}_\adot \bar{\xi}^\adot =: (\bar{\chi}\bar{\xi})\,.
\ee

In ${\mathrm{SL(2,}\,\mathbb{C})}$ notation, a four-component Dirac spinor is represented by a pair of chiral spinors:
\be
\Psi_{\mathrm{D}} = \left(\begin{array}{c} \psi_\alpha \cr \bar{\chi}^\adot \end{array}\right).
\ee
For a Majorana spinor, $\bar{\chi}_\adot = \overline{(\psi_\alpha)}$. The Dirac matrices are
\be
\Sigma^a = \left(\begin{array}{cc} 0 & (\sigma^a)_{\alpha\adot} \\ (\tilde{\sigma}^a)^{\adot\alpha} & 0 \end{array} \right),
\ee
where the matrices $\sigma^a$ ($a = 0,\ldots,3$) are defined as
\be
(\sigma^a)_{\alpha\adot} = (-{\mathbb{I}_2}, \vec{\sigma})_{\alpha\adot}\,, \qquad (\tilde{\sigma}^a)^{\adot\alpha} = \varepsilon^{\adot\bdot}\varepsilon^{\alpha\beta}(\sigma^a)_{\beta\bdot} = (-{\mathbb{I}_2}, -\vec{\sigma})^{\adot\alpha}\,,
\ee
with $\mathbb{I}_2$ the $2 \times 2$ identity matrix and $\vec{\sigma}$ the Pauli matrices
\be
\sigma^1 = 
\left(\begin{array}{cc} 0 & 1 \cr 1 & 0 \end{array}\right), \qquad \sigma^2 = \left(\begin{array}{cc} 0 & -i \cr i & 0 \end{array}\right), \qquad \sigma^3 = \left(\begin{array}{cc} 1 & 0 \cr 0 & -1 \end{array}\right),
\ee
and have the following properties:
\be
\begin{array}{l}
(\sigma^a)_{\alpha\adot}(\tilde{\sigma}_a)^{\bdot\beta} = -2\delta_\alpha^\beta \delta_\adot^\bdot\,, \qquad (\sigma_a)_{\alpha\adot}(\tilde{\sigma}^b)^{\adot\alpha} = -2\delta_a^b\,, \cr
\sigma^a\tilde{\sigma}^b = -\eta^{ab} +\sigma^{ab}\,, \qquad \tilde{\sigma}^a \sigma^b = -\eta^{ab} +\tilde{\sigma}^{ab}\,, \cr
\sigma^{ab} = -\sigma^{ba}\,, \qquad \tilde{\sigma}^{ab} = -\tilde{\sigma}^{ba}\,, \qquad (\sigma^{ab})_\alpha{}^\alpha = (\tilde{\sigma}^{ab})^\adot{}_\adot = 0\,,

\end{array}
\ee
with $\eta^{ab} = {\mathrm{diag}}(-1,1,1,1)$. These properties imply $\{\Sigma^a,\Sigma^b\}=-2\eta^{ab}\,\mathbb{I}_4$.

\subsection{Dimensional reduction}\label{dimred}

\

Since in the text we write expressions both in ten- and four-dimensional notation, it is important to clarify our notation and conventions. Breaking the SO($9, 1$) Lorentz symmetry to $\mathrm{SO}(3, 1) \times \mathrm{SO}(6)$ $\simeq$  $\mathrm{SO}(3, 1) \times \mathrm{SU}(4)$, an SO($9, 1$) vector $v^\mu$ ($\mu = 0, \ldots, 9$) decomposes as

\begin{equation}
v^\mu \;\longmapsto\; (v^a, v^{[ij]})\,,
\end{equation}
where $v^a$ ($a = 0,\ldots,3$) transforms under the representation {\bf 4} of SO($3, 1$) and $v^{[ij]} = -v^{[ji]}$ ($i,j = 1,\ldots,4$) transforms under the {\bf 6} of SU(4). The relation between the {\bf 6} of SU(4) and the {\bf 6} of SO(6) is given by the SO(6) Pauli matrices $(\rho_I)^{ij}=-(\rho_I)^{ji}$ ($I = 1,\ldots,6$) in the following way:
\be
v^{[ij]}=\frac{1}{2\mathrm{i}}(\rho_I)^{ij} v^{I+3}\,.
\ee
These matrices have the properties \cite{GSW}
\ba
(\rho^I)^{ij}(\rho^J)_{jk} +  (\rho^J)^{ij}(\rho^I)_{jk}   & = &   2\eta^{IJ}\delta_k^i \,, \nonumber \\
(\rho^I)_{ij}   & = &   \frac{1}{2}\varepsilon_{ijk\ell}(\rho^I)^{k\ell}\,, \\
(\rho^I)_{ij}(\rho_I)_{k\ell}   & = &   -2\varepsilon_{ijk\ell}\,, \nonumber
\ea
where $\eta^{IJ} = {\mathrm{diag}}(1,1,1,1,1,1)$ and $\varepsilon_{ijk\ell}$ is the SU(4)-invariant, totally antisymmetric tensor such that $\varepsilon_{1234}=1$. Analogously, one can define the tensor $\varepsilon^{ijk\ell}$ such that $\varepsilon^{1234}=1$. These satisfy the relation
\be
\varepsilon_{ijk\ell}\varepsilon^{k\ell mn} = 4\delta_{[i}^m \delta_{j]}^n\,.
\ee

A left-handed Majorana-Weyl spinor $\xi^\ahat$ ($\ahat = 1, \ldots, 16$) transforming under the  {\bf 16} of $\mathrm{SO}(9, 1)$ decomposes as

\be
\xi^\ahat \;\longmapsto\; (\xi^{\alpha i},\bar{\xi}_j^\adot)\,,
\ee
where we use the standard two-component notation for chiral spinors ($\alpha = 1,2\,; \adot = \dot{1},\dot{2}$) and $\xi^{\alpha i}$ (resp. $\bar{\xi}_j^\adot$) transforms under the representation {\bf 4} (resp. $\bar{\mathbf{4}}$) of SU(4). Analogous conventions apply to right-handed Majorana-Weyl spinors of $\mathrm{SO}(9, 1)$.

We also need to know how to translate the $\mathrm{SO}(9, 1)$ Pauli matrices $(\gamma^\mu)_{\ahat\bhat}$ and $(\gamma^\mu)^{\ahat\bhat}$ to the language of $\mathrm{SO}(3, 1) \times \mathrm{SU}(4)$. Based on \cite{mor}, we propose the following {\sl ansatz} for the non-vanishing components:

\ba
(\gamma^{a})_{(\alpha i) {\adot \choose j}}   & = &  \delta_i^j (\sigma^a)_{\alpha\adot} = (\gamma^{a})_{{\adot \choose j} (\alpha i)}\nonumber\\
(\gamma^{[k\ell]})_{(\alpha i) (\beta j)}   & = &  2\varepsilon_{\alpha\beta} \delta_{[i}^k \delta_{j]}^\ell \label{gammadown}\\
(\gamma^{[k\ell]})_{{\adot \choose i} {\bdot \choose j}}   & = &   \varepsilon_{\adot\bdot}\varepsilon^{ijk\ell}\nonumber
\ea
for $(\gamma^\mu)_{\ahat\bhat}$ and

\ba
(\gamma^{a})^{(\alpha i) {\adot \choose j}}   & = &  \delta_j^i (\tilde{\sigma}^a)^{\adot\alpha}  = (\gamma^{a})^{{\adot \choose j} (\alpha i)}\nonumber\\
(\gamma^{[k\ell]})^{(\alpha i) (\beta j)}   & = &  \varepsilon^{\alpha\beta}\varepsilon^{ijk\ell} \label{gammaup}\\
(\gamma^{[k\ell]})^{{\adot \choose i} {\bdot \choose j}}   & = &  2\varepsilon^{\adot\bdot} \delta_{[i}^k \delta_{j]}^\ell \nonumber
\ea
for $(\gamma^\mu)^{\ahat\bhat}$. It is straightforward to show that the above matrices satisfy the usual relation

\be
(\gamma^\mu)_{\ahat\bhat} (\gamma^\nu)^{\bhat\ghat} + (\gamma^\nu)_{\ahat\bhat} (\gamma^\mu)^{\bhat\ghat} = -2\eta^{\mu\nu}\delta_\ahat^\ghat\,,
\label{commutation}
\ee
with $\eta^{[ij][k\ell]} := \frac{1}{2} \varepsilon^{ijk\ell}$.

As an example, we show  how to obtain the dimensional reduction of the pure spinor constraints $\lambda\gamma^\mu\lambda=0$ using (\ref{gammadown}). For $\lambda\gamma^a\lambda=0$, we have
$$
\lambda^\ahat (\gamma^{a})_{\ahat\bhat} \lambda^\bhat = 0 \; \Longleftrightarrow \; \lambda^{\alpha i}(\gamma^{a})_{(\alpha i) {\adot \choose j}}\blambda^\adot_j + \blambda^\adot_j(\gamma^{a})_{{\adot \choose j} (\alpha i)}\lambda^{\alpha i} = 2\lambda^{\alpha i}(\sigma^a)_{\alpha\adot}\blambda^\adot_i = 0\,,
$$
whence
\be
\lambda^{\alpha i}\blambda^\adot_i = 0\,.
\ee
For $\lambda\gamma^{[ij]}\lambda=0$, we have
$$
\lambda^\ahat (\gamma^{[ij]})_{\ahat\bhat} \lambda^\bhat = 0 \; \Longleftrightarrow \; \lambda^{\alpha k} (\gamma^{[ij]})_{(\alpha k) (\beta \ell)}\lambda^{\beta \ell} + \blambda_k^\adot (\gamma^{[ij]})_{{\adot \choose k} {\bdot \choose \ell}} \blambda_\ell^\bdot = 2(\lambda^i\lambda^j)-\varepsilon^{ijk\ell}(\blambda_k\blambda_\ell) =0\,,
$$
whence
\be
(\lambda^i\lambda^j)=\frac{1}{2}\varepsilon^{ijk\ell}(\blambda_k\blambda_\ell)\,.
\ee

\pagebreak{}
\end{document}